# Photoluminescence in electronic ferroelectric $Er_{1-x}Yb_xFe_2O_4$

R. Wang, H. X. Yang[†], Y. B. Qin, B. Dong, J. Q. Li, and Jimin Zhao*

*Beijing National Laboratory of Condensed Matter Physics and Institute of Physics,*

*Chinese Academy of Sciences, Beijing 100190, China*

[†] *E-mail: hxyang@iphy.ac.cn*

*E-mail: jmzhao@iphy.ac.cn*

**ABSTRACT**

Strong Stark splitting, which is nearly independent of the R-ions replacement, has been observed through the photoluminescence investigation of electronic ferroelectric $Er_{1-x}Yb_xFe_2O_4$ ($x$=0, 0.8, 0.9 and 0.95). Initially multiple radiative decay channels have been investigated, especially the visible transition $^4F_{9/2} \rightarrow {}^4I_{15/2}$, of which a quenching effect has been observed. A series of small non-Raman peaks have been observed superimposed on a broadband photoluminescence spectrum, of which we tentatively assign Stark splitting to be the cause. The splitting of the $^4F_{9/2}$ and $^4I_{15/2}$ levels is found to be 54meV and 66meV, respectively. This unusually large Stark splitting at visible range indicates the existence of strong local field originated from the W-layer in the charge-frustrated $ErFe_2O_4$.

**INTRODUCTION**

The electronic ferroelectricity coupled with magnetism in charge-/spin- frustrated $RFe_2O_4$ (R=Y, Er, Yb, Tm and Lu) materials have attracted much attention in recent studies [1–5]. Experimental measurements in this kind of materials have revealed a notable interplay among charge, spin and the crystal lattice. Multiferroic properties [1], colossal dielectric constant [1], nonlinear current-voltage behavior, giant magneto-dielectric response [2] and colossal dielectric tunability [3] have been observed in a variety of well-characterized samples. These remarkable functionalities could turn out to be quite useful for developing novel devices for technologic applications.

From structural point of view, $RFe_2O_4$ has a layered rhombohedral structure with space group R-3m, this material consists of two structural layers, i.e. a hexagonal $Fe_2O_{2.5}$ double layer, which is made up of two triangular sheets of corner-sharing $FeO_5$ trigonal bipyramids and usually called W layer, sandwiched by a thick $R_2O_3$ layer (Fig. 1a). The origin of ferroelectricity in this kind of material is fundamentally correlated with the disappearance of the inversion symmetry as a consequence of the $Fe^{3+}$ and $Fe^{2+}$ charge ordering of W layer [1, 4–5]. Experimental studies revealed a large spontaneous polarization of about 0.24 $Cm^{-2}$ in $LuFe_2O_4$ [1], which is comparable to the polarization of 0.26 $Cm^{-2}$ in $BaTiO_3$ [6]

Erbium-doped ferroelectric materials have attracted much interest in recent years for their unique optoelectronic properties [7-9]. Doping magnetic Erbium ions into $La_2Ti_2O_7$ was attempted as one of the most promising procedures to realize noncentrosymmetric magnets, which manifests a unique nonreciprocal magneto-optical effect [9]. Ferroelectric $RFe_2O_4$ is a good host material, which is robust against $Er^{3+}$ substitution of the trivalent lanthanoid sites from a small amount up to a full replacement with its ferroelectricity remain polarized along the *c*-axis.

In this paper, we present a systematic photoluminescence (PL) investigation of the multiferroic $Er_{1-x}Yb_xFe_2O_4$ (*x*=0, 0.8, 0.9 and 0.95) materials. Due to its fluorescence nature, the Erbium ion is found to behave as an effective local probe of the crystal field or electric polarization induced by charge-frustration. We give a tentative interpretation of our observation, which may add confirmation to the existence of strong local field and/or electric polarization for these charge-frustrated materials.

**EXPERIMENT AND RESULTS**

We synthesized four polycrystalline samples $Er_{1-x}Yb_xFe_2O_4$ (*x*=0, 0.8, 0.9 and 0.95) using a conventional solid-state reaction method. A stoichiometric amount of $Er_2O_3$ (99.99%), $Yb_2O_3$ (99.99%) and guaranteed reagent grade $Fe_2O_3$ were fully mixed in an agate mortar. Then these raw materials were pressed into pellets and sintered at 1200 °C for 12 hours in an atmosphere, of which the oxygen partial pressure was controlled by adjusting a mixture of the gases $CO_2$ and $H_2$ ($CO_2/H_2 = 2$).

To characterize the crystal lattice structure and phase purity, powder X-ray diffraction (XRD) measurements were taken using a Rigaku diffractometer with the CuKα line. In Fig. 1(c) we show the XRD patterns taken from $Er_{1-x}Yb_xFe_2O_4$ (*x*=0, 0.8, 0.9 and 0.95) samples at room temperature, which can be confidently indexed using hexagonal cells with the R-3m space group. No peaks from impurity phases were observed. The relative intensity difference of the (009) peak for $x=0.9$ is very likely caused by large grain sizes and the preferred orientation effect. Our scanning electron microscopy characterization revealed that the composing crystallines for each of the four as-made materials demonstrate typical layered feature with crystal sizes of 1-10μm. From these XRD measurements we obtained, using a software PowderX, the lattice constants, which show weak dependence on doping concentration and are listed in Table 1.

Then transmission electron microscopy measurements were taken on a Tecnai F20 (200 kV) electron microscope, clearly confirming very similar superstructure features as observed in $LuFe_2O_4$ [4-5]. Particularly, we show in Fig. 1(b) a typical room-temperature electron diffraction pattern of $ErFe_2O_4$ taken along the direction of $[1\bar{1}0]$ zone-axis. Similar to that commonly accepted for $LuFe_2O_4$, the superstructure reflections, as diffuse streaks on the (h = 3, h =3, l) lines along the $c$-axis, arise essentially from the charge ordering transition associated with the mixed valence states of Fe ions in the frustrated layers[1, 4-5], which correlates with the electronic ferroelectricity.

To do the PL measurements, the four millimeter-sized samples were carefully polished such that their surfaces are flat and optically smooth, except for noticeable dents on the surfaces due to their polycrystalline nature. Their roughness is different and the $x=0.8$ one has much more dents on the surface than the other three samples have.

We then measured the room-temperature fluorescence of the four $Er_{1-x}Yb_xFe_2O_4$ ($x=0$, 0.8, 0.9 and 0.95) samples with a confocal microscope Raman-fluorescence spectrometer system, where the excitation continuous wave laser beam has a wavelength of 532nm and powers tuned from 0.02mW to 4mW. The fluorescence light was collected with the same objective lens for excitation, frequency-resolved with the spectrometer grating, and then detected with the charge-coupled device (CCD) of the spectrometer. The integration time was 10s for each of the measurements and the spectra are shown in Fig. 2. When different spots of the sample surface were sampled, different fluorescence intensities were observed, and when the sample surface roughness is high, the fluorescence intensity is relatively low. We collected signals at relatively smooth parts of the surfaces.

The power dependence of the fluorescence for the four samples is shown in Fig. 2. It can be seen that all the emission curves for all the samples have a visible broad-band spectrum range, which does not vary noticeably with different excitation powers. Unlike this, the spectrum intensity varies with the excitation power and doping concentrations of the $Yb^{3+}$ ions. To illustrate explicitly, we show the integrated fluorescence intensity versus the entire excitation power in Fig. 3. It is known that, for unsaturated processes, the number of photons needed for populating an excited state can be obtain according to the following relation [10]

$$I \propto P^n \quad (1),$$

where $I$ is the fluorescence intensity, $P$ is the excitation power, and $n$ is the number of pumping photons needed for the quantum transition from the ground state to the emission state. As shown in Fig. 3, the $n$ values are very close to 1 for $x=0$, 0.8 and 0.9, but larger than 1 for $x=0.95$. Thus for relatively lower $Yb^{3+}$ concentration ($x \leq 0.9$), the excitation is dominated by one-photon processes, but for higher $Yb^{3+}$ concentration ($x=0.95$) the excitation is a combination of one-photon and two-photon processes. For the latter case, one-photon process is still dominating even if the $Er^{3+}$ concentration becomes as small as

0.05. Overall speaking, linear fitting using eq. (1) well describes this down-conversion process [11].

Also can be seen from Fig. 3 is the decrease of slop at higher excitation powers, which is the so-called quenching effect [10]. We ascribe this to the change of decay channels at high powers, where more decay is through channel ① rather than channel ② (channel ① and ② are marked in Fig. 4). Shown in Fig. 2, more emission is observed with a wavelength corresponding to channel ① (around 546 nm) at high powers (for example, 4mW excitation power), confirming our above analysis. This quenching effect becomes less observable with increasing $Yb^{3+}$ concentration. Thus intense doping of $Yb^{3+}$ ions helps maintain or enhance illumination at the 600-800nm range, *i.e.* that through channel ②. This is possible partially through channel ④ and ⑤ (Fig. 4).

To identify the PL channels, we explicitly illustrate the energy levels of the 4f electronic states of doping $Er^{3+}$ and $Yb^{3+}$ ions in Fig. 4. Under the excitation of 532 nm photons, the ground state of $Er^{3+}$ ($^4I_{15/2}$) is excited to the higher energy state ($^4S_{3/2}$), with associated multi-phonon processes that help to preserve energy conservation. Then the excited state ($^4S_{3/2}$) decay radiatively $^4S_{3/2} \rightarrow {}^4I_{15/2}$ (546nm) to the ground state ($^4I_{15/2}$) and non-radiatively to the lower levels ($^4F_{9/2}$, $^4I_{9/2}$, $^4I_{11/2}$ and $^4I_{13/2}$), followed by subsequent radiative transitions $^4F_{9/2} \rightarrow {}^4I_{15/2}$ (662 nm) and $^4I_{9/2} \rightarrow {}^4I_{15/2}$ (800 nm). Besides these one-photon processes, two-photon processes might also occur for very high $Yb^{3+}$ concentration (see the line for *x*=0.95 in Fig. 3). Two possible channels, marked by ④ and ⑤, are illustrated in Fig. 4. In such processes ④' and ⑤' are assisting transitions which partially comes from the phonon-assisted energy transfer from $Yb^{3+}$ ions to $Er^{3+}$ ions. When the $Yb^{3+}$ concentration is higher (e.g. $x=0.95$), the contribution from energy transfer is higher accordingly, leading to an $n$ value larger than 1.

It can be seen from Fig. 2 that the visible PL has a red shift for $x=0.95$ and 0.9, compared with that for $x=0$ and 0.8. When $Yb^{3+}$ ion has a high concentration, two-photon processes become easier, resulting in fast depletion of the $^4I_{11/2}$ and $^4I_{13/2}$ states. Thus phonon-assisted non-radiative transitions $^4S_{3/2} \rightarrow ^4F_{9/2}$, $^4F_{9/2} \rightarrow ^4I_{9/2}$, $^4I_{9/2} \rightarrow ^4I_{11/2}$, and $^4I_{11/2} \rightarrow ^4I_{13/2}$ are easier to occur. Consequently radiative transition channel ① is reduced and radiative transition channel ② has a relatively larger reduction in higher energy sub-channels. The net result is a red shift in the visible PL spectrum in Fig. 2 with increasing $x$. Meanwhile a careful examination of the bandwidth of the visible PL corresponding to channel ② does not give a noticeable broadening (make sure channel ① is not included here). Thus the red shift is unlikely due to Stark splitting (see the analysis section) at a higher $Er^{3+}$ ion concentration of 10% and 5%.

**ANALYSIS**

In order to get deeper insight into the PL phenomena, we re-plot in Fig. 5a a typical data (the light-blue curve in Fig. 2d) obtained from the $x=0.95$ compound with 0.4mW excitation power. Detailed examination of the spectrum revealed that there are multiple small peaks superimposed on the broad-band PL curve (the same to the other curves in Fig. 2). To determine the peak positions we subtracted, in energy domain, the spectrum curve with a Gaussian fitting curve and multiplied the residue values by 3 for figure clarity. The peak positions are indicated in Fig. 5(a). We first identify the radiative channels for the PL spectrum. While the major broadband peak centered at 1.8eV corresponds to radiative channel ②, the two bumps with energies around 2.1eV and 1.5eV apparently corresponds to the radiative channels ① and ③. Next we focus on the small peaks superimposed on the profile. The peaks with higher energies (2.1eV ≤ $E$ ≤ 2.3eV) are assigned to first- and second-order Raman active modes, which were confirmed by our varying-excitation-wavelength PL measurements (not shown here). Note that the starting value of the *x*-axis of Fig. 5(a) was set to be the excitation laser energy (2.3367eV, *i.e.* 532nm).

We further analyze the series of small peaks with relatively lower energies (1.56eV ≤ $E$ ≤ 2.02eV). We clearly identify the peak positions and find they are separated by two energy intervals 54meV (between blue lines) and 66meV (between magenta lines). We contemplate their origins and tentatively assign them to strong stark-splitting of the 4f electronic states of $Er^{3+}$ ions. Under stark splitting due to the broken symmetry of local field, the ground state $^4I_{15/2}$ splits into 8 sub-levels, and the excited-state $^4F_{9/2}$ splits into 5 sub-levels, as shown in Fig. 5(b). The possible combination of radiative transitions is depicted by arrows as presented in Fig. 5(b). The total transition intensity $I = \sum I_i$, where $I_i$ corresponds to the transitions shown in Fig. 5(b) and marked in Fig. 5(a), fits very well with the original experimental result. Such data analysis gives a fixed energy assignment and a nearly unique combination of transitions with evenly spaced splitting, which is unlike former treatments for amorphous glasses [12-14] or doped crystalline materials [11, 15], of which the energy assignment and number of transitions is not unique and the sub-levels are often not evenly-spaced in energy.

Most of former reports on PL of $Er^{3+}$ ions in amorphous [12-14] and crystalline [11, 15] materials were explained by Stark splitting, of which narrow green (around 548nm) and red

(around 660nm) bands were frequently observed, with an energy interval of typically 5-10 meV for the visible range. What we have observed for $Er_{1-x}Yb_xFe_2O_4$ (Fig. 5(b)) gives an energy interval of 54 meV and 66 meV, which is several times larger. We interpret this as due to the strong local field which is characteristic of ferroelectric materials. It is well known that $LuFe_2O_4$ has a layered lattice structure, and its spontaneous electric polarization due to charge-frustration is prominent with a direction along the $c$-axis [1, 4]. Thus, based on our analysis, our PL measurement confirms the existence of strong local field in the $RFe_2O_4$ multiferroic materials.

To investigate the effect of doping level to the magnitude of Stark splitting, similar analysis has been done also for the $x$=0, P=0.4mW case. In Fig. 6 we do not see noticeable broadening of the splitting. This further confirms our decay-channel analysis that PL corresponding to channel ② is not broadened. Thus we conclude that R-ion replacement does not noticeably affect the Stark splitting, which can be understood as the electric polarization is mainly contributed from the W-layer rather than the $R_2O_3$ layer.

**CONCLUSIONS**

In summary, we have systematically studied the photoluminescence of multiferroic $Er_{1-x}Yb_xFe_2O_4$ ($x$=0, 0.8, 0.9 and 0.95) materials. Power- and concentration-dependence measurements revealed that one-photon transition $^4F_{9/2} \rightarrow {}^4I_{15/2}$ (662 nm) is the major radiative channel, along with a quenching effect which is due to the competition between decay channels. The effect of co-doped $Yb^{3+}$ ions is also analyzed, which mainly contributes to two-photon transition processes. More provocatively, we have observed fine features of the emission spectra. While it is still open to further investigations, we tentatively assign the non-Raman peaks to strong Stark splitting, which can be explained by and also confirms the existence of strong local field within these multiferroic materials. With this optical method, we further show that the Stark splitting is due to the charge frustration of the Fe ions in the W-layer and is practically immune to lattice modification induced by rare-earth ion replacement in the $R_2O_3$ layer.

## ACKNOWLEDGEMENT

We thank H. X. Xu, Y. L. Liu and Yuhei Miyauchi for helpful discussions. This work is financially supported by the National Natural Science Foundation of China (10974246, 10704085, 10874227 and 10874214), the National Basic Research Program of China (2007CB936804).

**FIGURE AND TABLE CAPTIONS**

Figure 1. (a) Schematic unit cell structure depicting the Er–O layer, Fe–O double- and single-layers stacking alternately along the $c$-axis for the $ErFe_2O_4$ phase, with O-atoms omitted for clarity. (b) Electron diffraction patterns of $ErFe_2O_4$ along $[1\bar{1}0]$ zone-axis directions at room temperature. (c) The XRD patterns for $Er_{1-x}Yb_xFe_2O_4$ ($x$=0, 0.8, 0.9 and 0.95) samples.

Figure 2. Fluorescence spectra of the $Er_{1-x}Yb_xFe_2O_4$ ($x$=0, 0.8, 0.9, 0.95) samples as a function of the excitation power.

Figure 3. Integrated $^4F_{9/2} \rightarrow {}^4I_{15/2}$ fluorescence intensity as a function of the excitation power for $Er_{1-x}Yb_xFe_2O_4$ samples of different doping values. The solid lines are linear fits. The dashed lines are guides to the eye.

Figure 4. Schematic energy levels of $Er^{3+}$ and $Yb^{3+}$ ions in $Er_{1-x}Yb_xFe_2O_4$ showing

multi-phonon assisted radiative decays in the visible and near 800nm range.

Figure 5. Schematic Stark-splitting levels and corresponding transitions. (a) The red curve is the original experimental result and the black curve is the residual after subtraction of a Gaussian fit (multiplied by 3 for figure clarity). The magenta and blue vertical lines are guides to the eye, indicating two series of energy intervals. (b) Evenly-spaced stark splitting.

Figure 6. Stark-splitting analysis for the $x$=0, P=0.4mW case, in comparison with that shown in Fig. 5.

Table 1 . Crystal lattices for $Er_{1-x}Yb_xFe_2O_4$, where $x$ is the doping level.

**TABLE 1**

| $x$ | $A$(Å) | $c$(Å) |
|---|---|---|
| 0 | 3.49473 | 24.92979 |
| 0.8 | 3.46304 | 25.04790 |
| 0.9 | 3.46173 | 25.08053 |
| 0.95 | 3.45885 | 25.08109 |

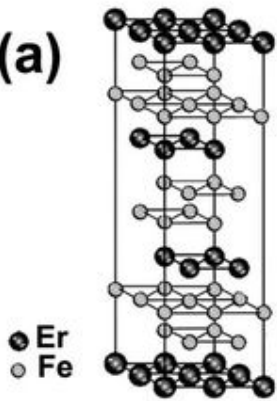 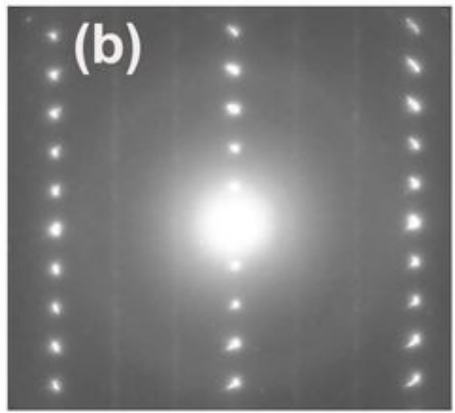 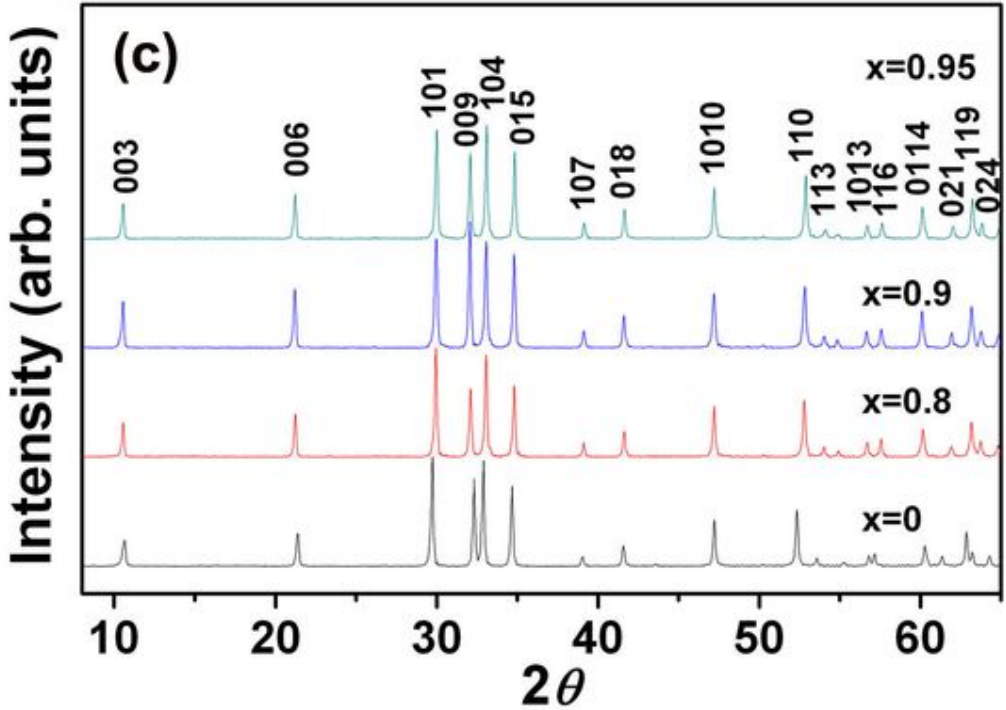

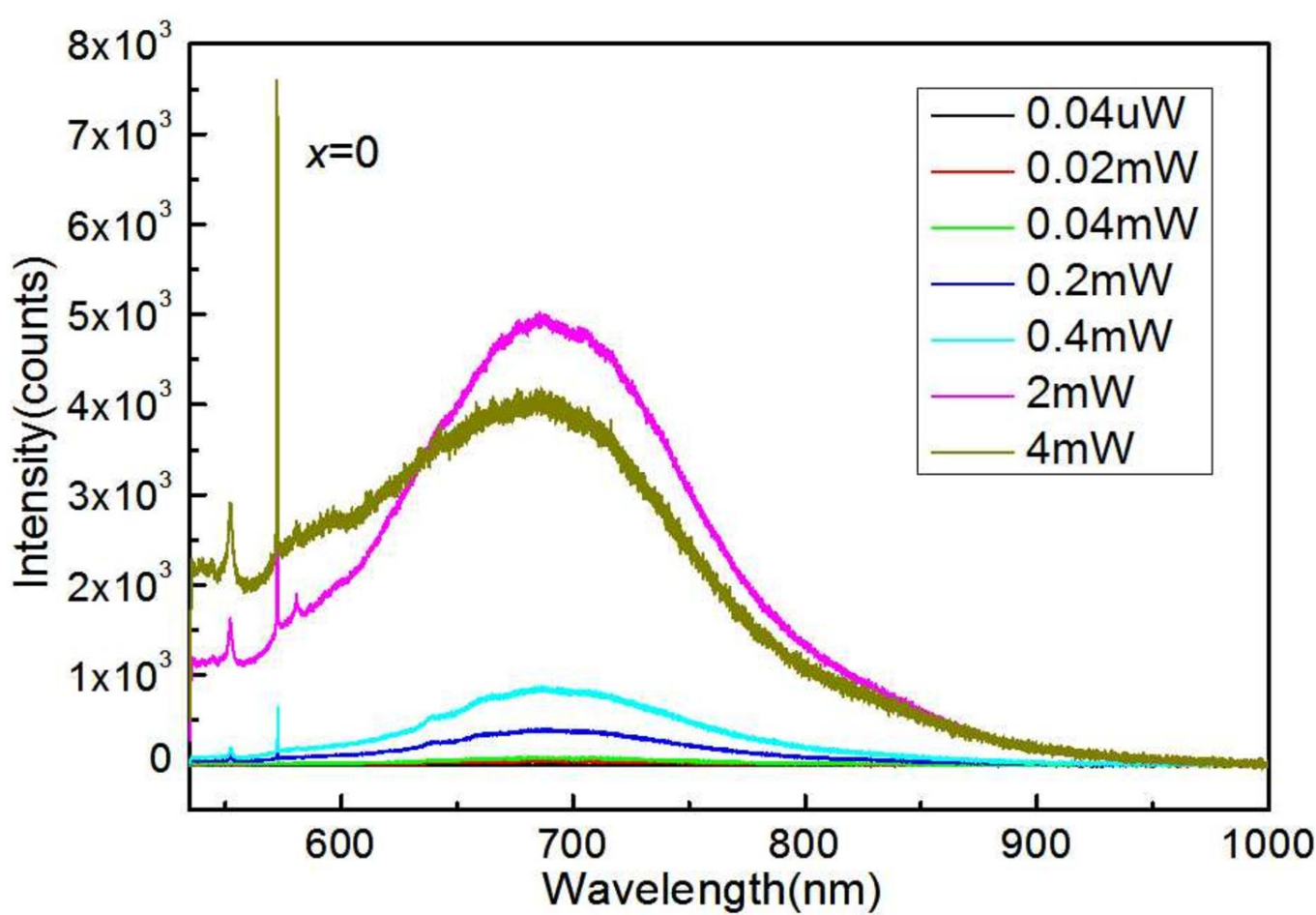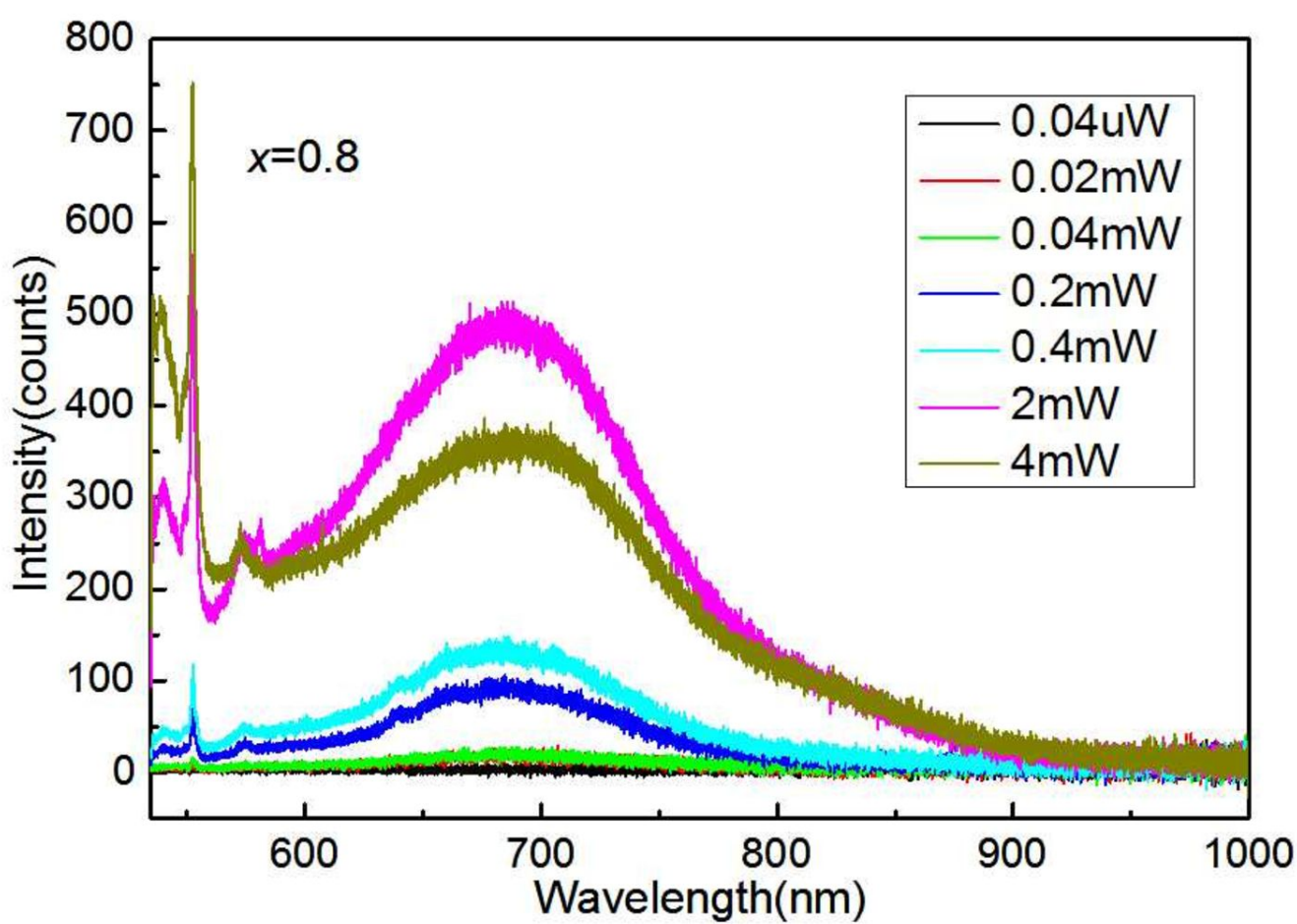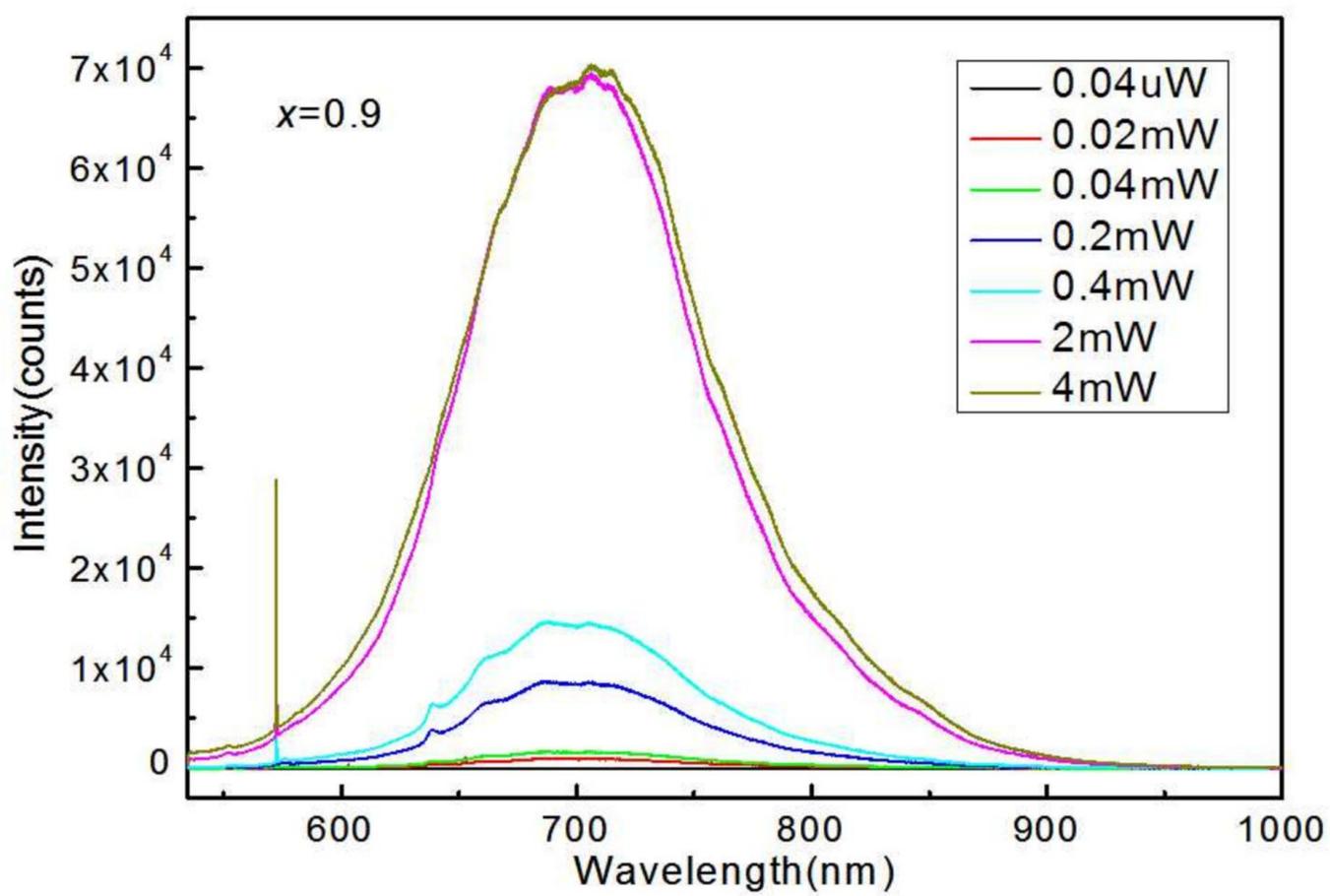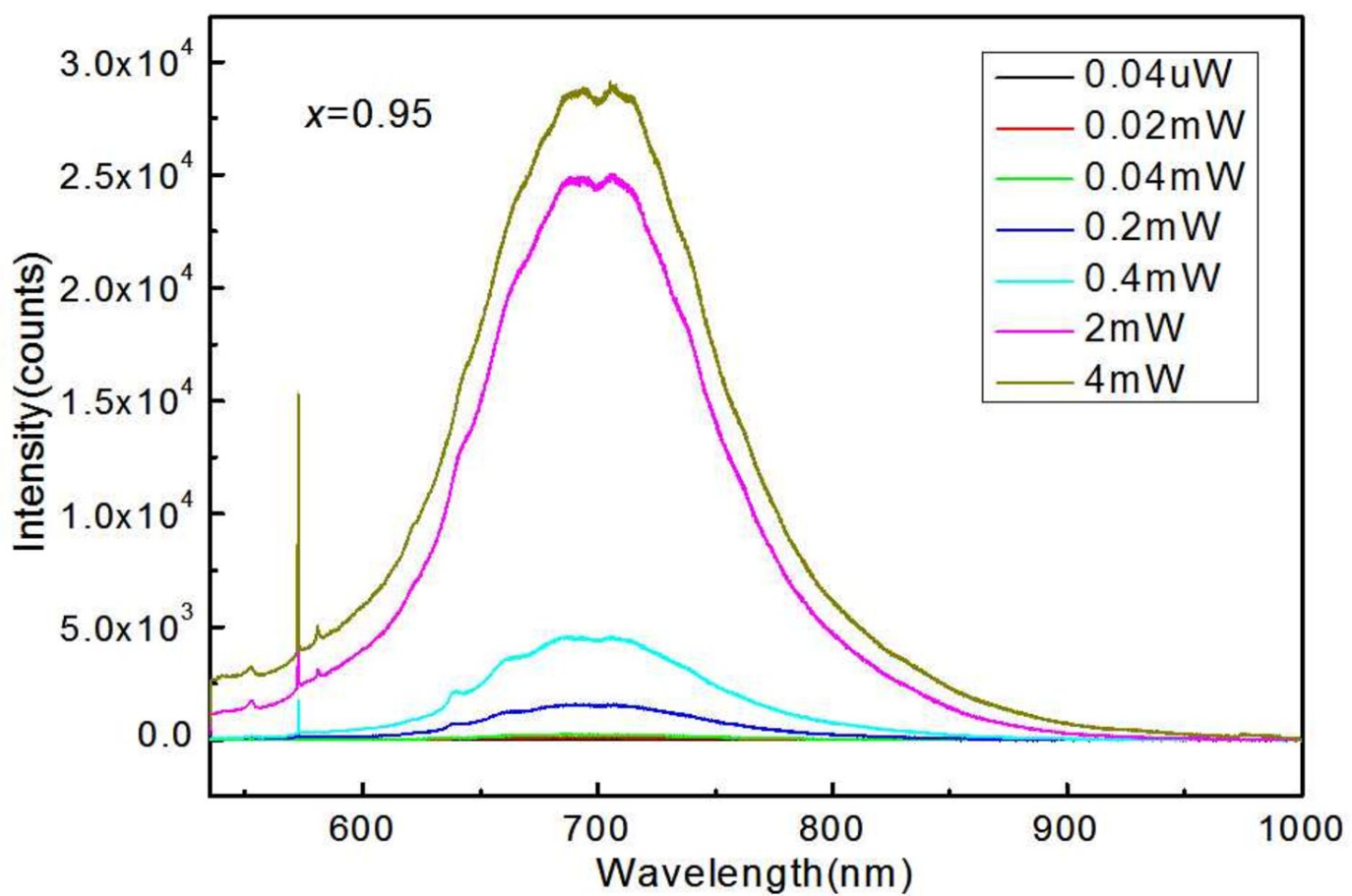

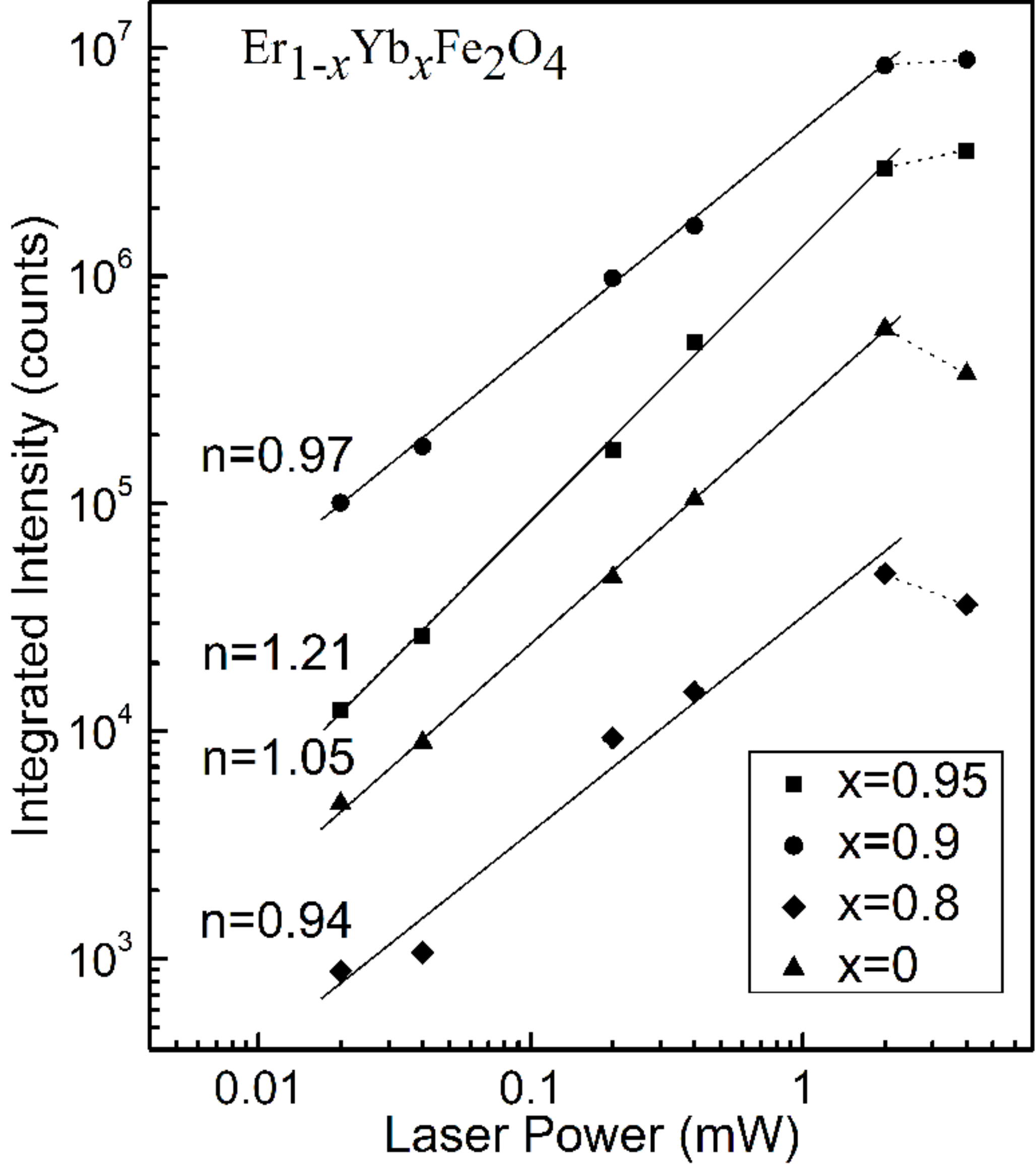

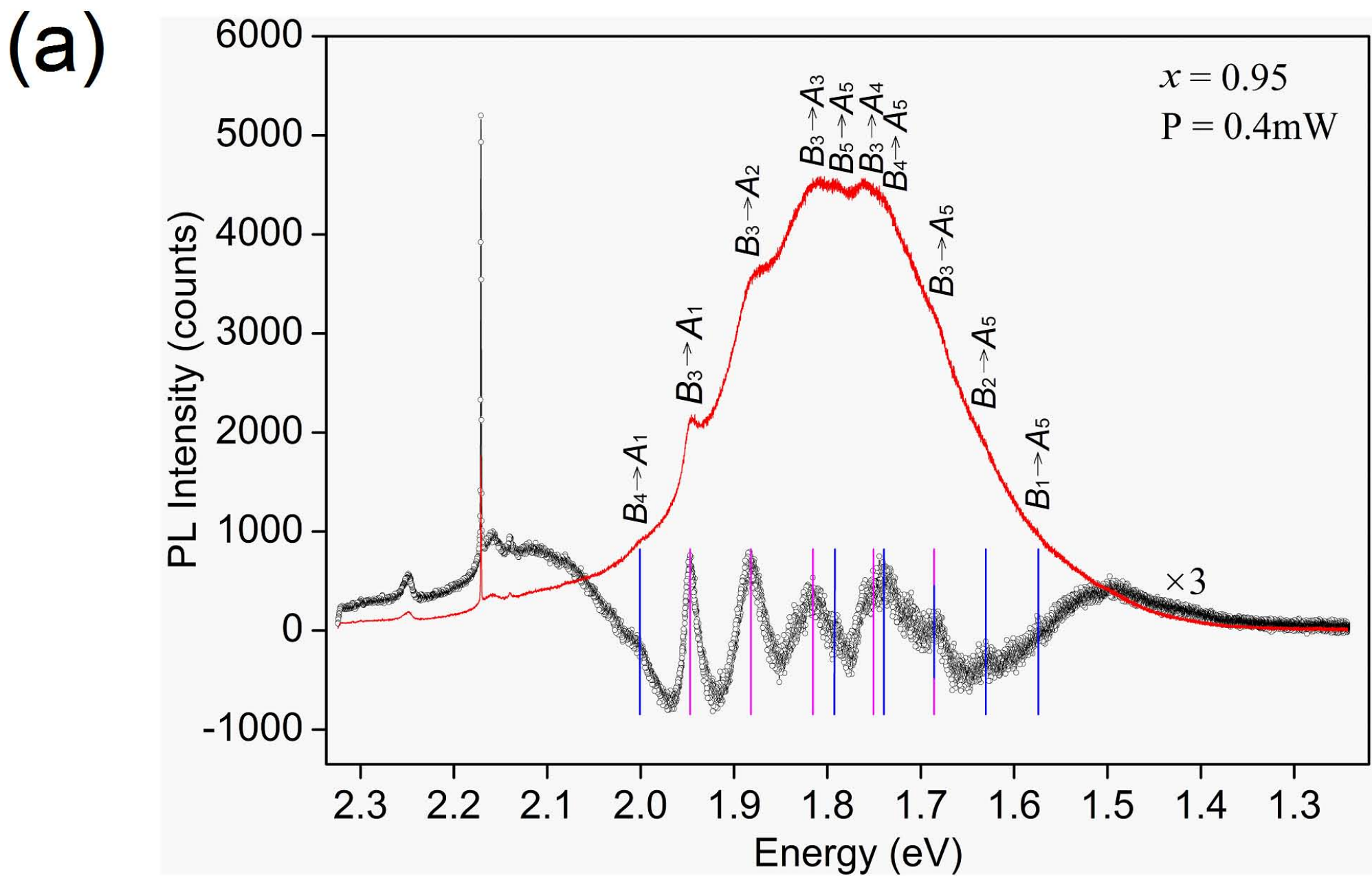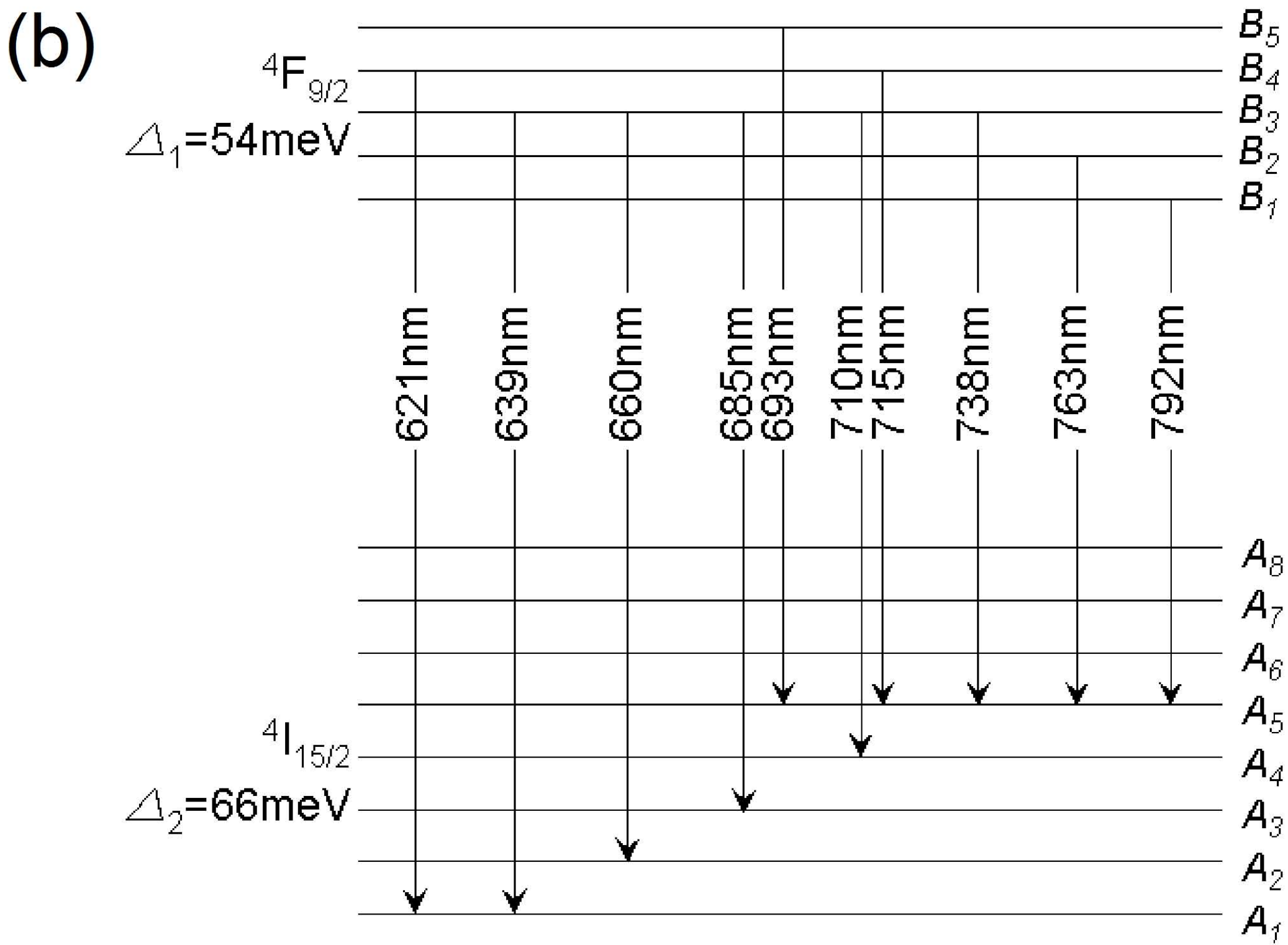

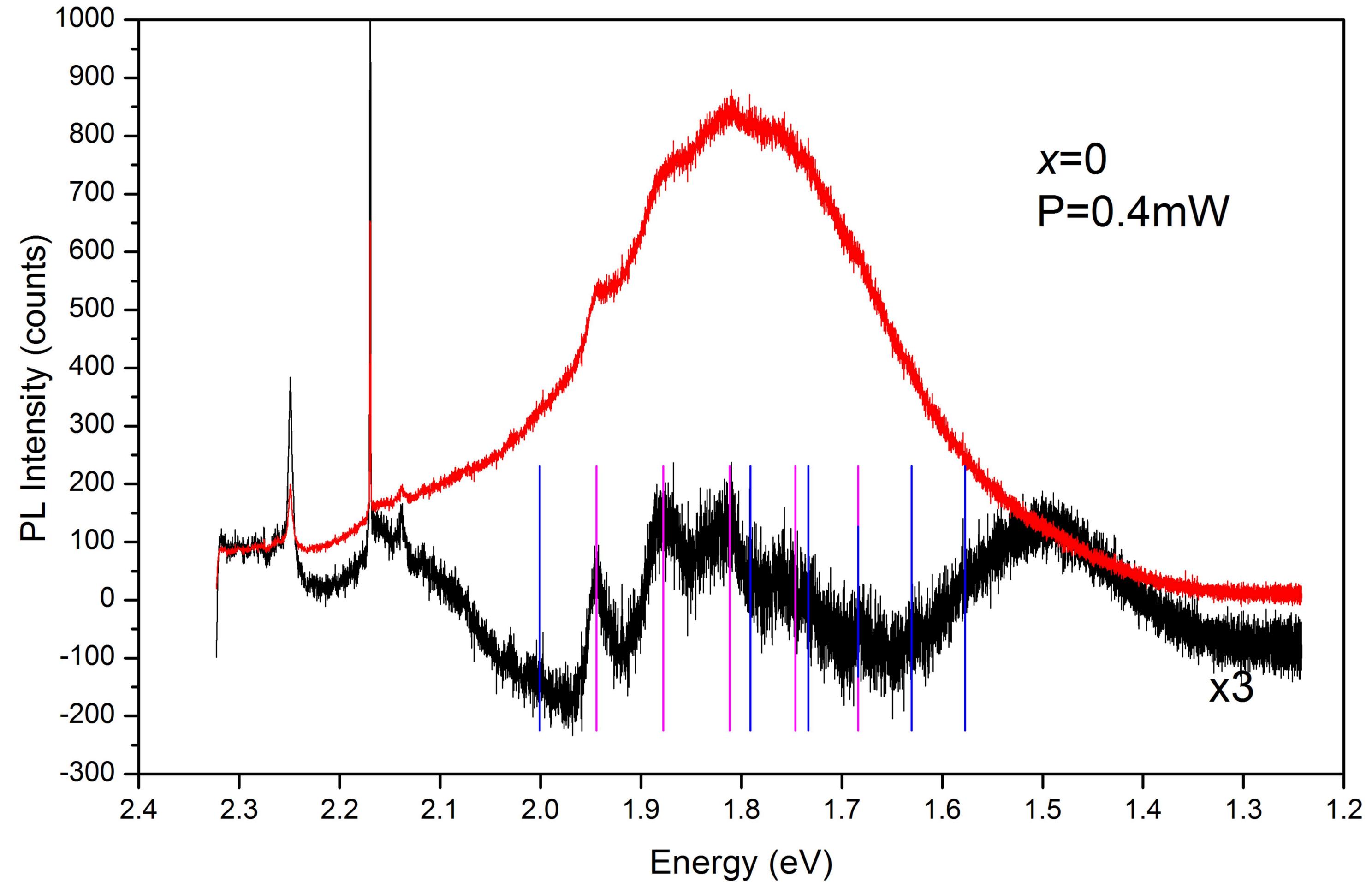